\documentstyle[aps,preprint]{revtex}
\begin{document}
\draft
\title{The Unusual Universality of Branching Interfaces in Random Media}
\author{Mehran Kardar$^{1}$, Attilio L. Stella$^2$, Giovanni Sartoni$^{3}$, and
 Bernard Derrida$^{4,5}$}
\address{$^1$ Department of Physics,
Massachusetts Institute of Technology,
Cambridge, Massachusetts 02139}

\address{$^2$  Dipartimento di Fisica and Sezione INFN, Universita' di Padova,
35131 Padova, Italy }

\address{$^3$ Dipartimento di Fisica and Sezione INFN, Universita' di Bologna,
40126 Bologna, Italy}

\address{$^4$ Laboratoire de Physique Statistique, Ecole Normale Sup\'erieure,
24 rue Lhomond, F-75005 Paris, France}

\address{$^5$ Service de Physique Th\'eorique, Centre d' Etudes  de 
Saclay, F-91191 Gif-sur-Yvette, France}
\date{\today}
\maketitle
\begin{abstract}
We study the criticality of a Potts interface
by introducing a {\it froth} model which,
unlike its SOS Ising counterpart, incorporates bubbles of
different phases. The interface is fractal at the
phase transition of a pure system. However, a position
space approximation suggests that the probability of loop 
formation vanishes marginally at a transition dominated by
{\it strong random bond disorder}. This implies a linear 
critical interface, and provides a mechanism for the 
conjectured equivalence of critical random Potts and
Ising models.
\end{abstract}
%%
%\begin{multicols}{2}  \narrowtext
%%
The effect of quenched impurities on phase transitions is 
important and quite fascinating.  A simple ``Harris 
criterion"\cite{Harris} indicates that critical behavior 
is modified by (bond) randomness in systems with a positive 
heat capacity exponent. Impurities can also change first 
order transitions to second order\cite{Berker}, to the 
extent that in two dimensions there are no discontinuous
phase transitions\cite{Aizenman}.  
There is a growing body of numerical\cite{Chen} and 
experimental\cite{Experiment} evidence that, at least in 
some situations, the asymptotic criticality is similar to 
the random bond Ising model, {\it irrespective of
the underlying symmtery}. Here we provide some 
justification for this observation based on an unexpected 
universality of the critical interface in the presence of
strong bond randomness.

Interfaces are key to second order phase transitions; the 
interfacial free energy vanishes at criticality with the 
Widom exponent of $\mu$. The behavior of Ising interfaces 
has been extensively studied by a solid--on--solid 
(SOS) model which simplifies numerical and theoretical 
analysis\cite{Forgacs}. Potts models also have a potential 
continuous transition with broken discrete symmetry\cite{Wu}. 
Potts interfaces are harder to study as they are 
complicated branching objects which become fractal at 
criticality. We introduce several approximations to such 
interfaces in 1+1 dimensions, eventually arriving at a 
model that is amenable to a position space renormalization 
group (RG) treatment; exact on a hierarchical lattice. The 
simplified model also allows us to examine interfaces of 
random bond Potts models. This leads to an intriguing 
generalization of directed polymers in random media\cite{DP}, 
with potential applications beyond those discussed here.
 
Consider the interface between two distinct phases of an 
ordered $q$ state Potts model. At very low temperatures the 
interface is a weakly fluctuating surface, much as in the 
Ising case. However, on approaching the critical point, 
bubbles of any of the other $q-2$ phases may appear at the 
interface, in addition to islands in the bulk phases. 
As in SOS models of Ising interfaces, we shall ignore 
isolated islands and overhangs. The resulting 
{\it froth} is a collection of bubbles, each bounded by two 
SOS surfaces. Even this simplified model is too hard to 
analyze: The allowed configurations in 1+1 dimensions
are a subset of those encountered in directed percolation 
(no dangling branches), with weights depending on $q$.
To make the problem tractable, we confine the interface to  
the bonds of a diamond hierarchical lattice (DHL) 
(inset of Fig.(1)).  In one iteration on this lattice
the interface can cross either of two branches, or create a
bubble by going through both. The bubble has a 
``fugacity" of  $q-2$, the number of possible intermediate 
phases. This procedure is repeated iteratively creating the 
possibility of loops within loops ad infinitum. Not all 
configurations of the original froth model (those with 
multiplicities that are not powers of $q-2$)
are included within this scheme. It is possible to  
construct more complicated models and RG schemes without 
these deficiencies; for example by considering the diamond 
plus diagonal hierarchical lattice (DDHL) in Fig.(1).
We have checked that the qualitative results are unchanged 
by the choice of RG scheme, and will thus focus on the 
simpler DHL where the recursion relation for the interface 
partition function is
\begin{equation}\label{PureRR}
Z_{n+1}=2Z_n^2+(q-2)Z_n^4\quad.
\end{equation}
Here $Z_1=\exp(-2\beta J)$ is the Boltzmann weight of one 
broken bond and $\beta=1/(k_BT)$.
 
The above recursion relation has a stable fixed
point at $Z^*=0$. For initial values flowing to this point 
$Z_n\propto\exp[-L f_s]$,  where $L=2^n$ is the length of 
the lattice after $n$ iterations and $f_s$ is the 
{\it interfacial free energy}. There is another stable sink 
with $Z^*\to\infty$ where $Z_n\propto\exp[-L^2 f_b]$. 
This phase is a dense foam of bubbles where the interface 
analogy  breaks down. These two phases are separated 
by an unstable fixed point at a {\it  finite} $Z^*(q)$.
Clearly any finite fixed point for $Z$ corresponds to 
$f_s=f_b=0$! A similar fixed point mechanism for the 
vanishing of the surface energy is present in wetting
phenomena\cite{wetting}. Denoting $t=Z_1-Z^*(q)$, we 
find $t'=2^{y(q)}t$ with $y(q)=2+\ln(1-Z^*(q))/\ln 2$. 
Since the interface free energy satisfies the homogeneity 
condition $f_s(t)=b^{-1}f_s(b^{y(q)}t)$, it vanishes as 
$t^\mu$ with $\mu(q)=1/y(q)$. The exponents obtained from
the DHL and DDHL RG schemes are compared with the exactly 
known values for the Potts model in $d=2$ in table I.
The exponent is (accidentally) exact for $q=2$ on the DHL, 
and shows the correct trend on increasing $q$. However,
this approach does not show the expected change  
to first order transitions at $q=4$\cite{Wu}. The critical 
interface is  {\it fractal}, its mass (number of occupied 
bonds) growing as $M=d\ln Z/d\ln Z_1\propto L^{d_f}$ with 
$d_f=y(q)$. The probability of forming a loop is 
{\it independent} of scale and given by $P=2^{d_f-1}-1$. 
%The hyperscaling expression $\mu=(d-1)\nu=\nu$, 
%relates the divergence of the correlation length $\xi$
%to the vanishing of $f_s$. {\it In two dimensions} we 
%can exploit the duality of the high temperature expansion 
%for the two point correlation function with the above low 
%temperature expansion of the interface energy. The set of 
%graphs are similar, with $\xi^{-1}$ replacing $f_s$. 
%This again  leads to a divergence, $\xi\propto t^{-1/y(q)}$.
 
As shown by Hui and Berker\cite{Berker}, randomness in the  
couplings of a Potts model may change first order 
transitions to second order. In particular, there can be 
no discontinuous symmetry breaking in {\it two 
dimensions}\cite{Berker}\cite{Aizenman}, and all disordered 
Potts models have continuous phase transitions. This opens 
up the possibility of an infinite set of new universality  
classes. A numerical study\cite{Chen} of the 8--state Potts 
model has confirmed the second order nature of the transition.  
Intriguingly, based on their simulations, the authors of 
ref.\cite{Chen} conjecture that all these new universality 
classes are in fact similar to the random bond Ising ($q=2$) 
model. There is also some experimental support for this 
conjecture from the effects of oxygen impurities on the 
phase transition of $(2\times 2)-2${\tt H} on {\tt Ni}(111). 
The pure system is a realization of the $q=4$ Potts model, 
while in the presence of impurities the asymptotic behavior 
appears to be Ising--like\cite{Experiment}.
This contradicts other evidence based on RG
schemes\cite{RBRG}\cite{Ludwig} that, at least for weak 
disorder, the critical exponents do depend on $q$. In fact
the latter RG scheme\cite{Ludwig} was partly motivated by an
earlier work\cite{Dotsenko} suggesting that the (nonuniversal)
critical line of the pure Baxter model shows Ising behavior 
in the presence of randomness.
 
To gain further insight into random bond criticality, 
we examine the singular behavior of the interfacial energy
in the framework developed earlier. The initial 
weights $\{Z_1(i)\}$ for the bonds of the DHL are chosen 
randomly from a probability distribution ${\cal P}_1(Z)$. 
The recursion relation for a specific realization of bonds,
\begin{eqnarray}\label{ZRR}
Z_{n+1}&&=Z_n(1)Z_n(2)+Z_n(3)Z_n(4) \nonumber\\
&&+(q-2)Z_n(1)Z_n(2)Z_n(3)Z_n(4) ,
\end{eqnarray}
can be used to construct a functional recursion relation 
for ${\cal P}_n(Z)$. We use a binary initial distribution 
of energies with a fraction $p$ of positive bonds $J_1$, 
and $1-p$ of positive {\it or negative} bonds $J_2$, i.e.
\begin{equation}\label{Pone}
{\cal P}_1(Z)= p\delta\left(Z-e^{-2\beta J_1}\right)+
(1-p)\delta\left(Z-e^{-2\beta J_2}\right).
\end{equation}
Under iteration, the simple fixed points of the pure system 
are replaced by stable distributions: The analog of $Z^*=0$ 
describes directed  paths on the hierarchical lattice and 
is governed by the distribution discussed in ref.\cite{Derrida}. 
This distribution for $\ln Z$ is characterized by a mean 
that shifts as $-f_sL\ll 0$ and a width that grows as 
$L^{\omega}$ with $\omega\approx 0.30$ (compared to the exact 
value of $\omega=1/3$ on a two dimensional lattice\cite{DP}). 
In the dense foam phase with $\overline{\ln Z}=-f_b L^2\gg 0$,
the central limit theorem should apply and we expect 
fluctuations of $\ln Z$ to grow as $L$. An argument similar 
to the Harris criterion\cite{Harris} shows that the 
relevance of randomness at the critical fixed point is 
determined by the sign of $y(q)-1$. Thus any randomness is 
expected to modify the interfacial criticality of Potts 
models with $q\geq 2$. By analogy with the other two 
limiting distributions, we expect the critical point to 
flow towards a third zero temperature stable distribution 
with $\overline{\ln Z}=0$, and $\delta\ln Z\propto L^{\omega_c}$.
 
These expectations are consistent with numerical iterations 
of the recursion relation starting from a large initial 
ensemble of $\{Z_1(i)\}$\cite{Dist}. The RG flows are 
towards zero temperature, leading to the phase diagrams 
indicated in Fig.(1) for $J_2=-J_1$ (and $J_2=+J_1/2$) 
for $q=3$ and $4$. The choice of an initial distribution 
with a fraction of {\it negative bonds} ensures that the 
phase boundary extends to zero temperature\cite{Dilution}. 
Since the eventual fixed distributions are at zero 
temperature, criticality can be examined by directly
looking at the recursion relations at $T=0$. With this 
enormous simplification, we have iterated energies,
\begin{eqnarray}\label{ERR}
E_{n+1}&&=\min\{E_n(1)+E_n(2),E_n(3)+E_n(4), \nonumber\\
&&E_n(1)+E_n(2)+E_n(3)+E_n(4)\} .
\end{eqnarray}
The last term is of course absent for $q=2$. We could 
iterate exactly the probability distribution for bond energies
up to $n=7$. The Monte Carlo iteration was typically extended 
up to $n=20$ for our determinations of the $T=0$ critical 
properties. It is not immediately apparent from Fig.(1) that 
the phase boundary for an initial mixture of ferromagnetic 
bonds is also be governed by a $T=0$ fixed point that has 
a mixture of positive and negative bonds. Although in a 
regular RG scheme positive bonds only generate positive 
bonds, it can be checked easily that for all temperatures 
intermediate between the extreme critical points, the 
recursion of eq.(2) leads to a mixture of positive and 
negative bonds. Thus the critical behavior is likely to 
be the same for both cases.
 
A striking feature of the $T=0$ recursion relation is 
that it is {\it independent of $q$} for $q\neq 2$. Thus 
the critical behavior for the vanishing of $f_s$ should 
not depend on $q$. This expectation goes beyond the 
approximations of the model and the hierarchical lattice. 
Consider the configurations contributing to a low temperature 
expansion of the interface free energy of the Potts 
model on any lattice. The index $q$ appears only in 
entropic factors giving the multiplicity of possible 
colorings. If, as is usually the case in random systems, 
the scaling properties of the interface are controlled 
by a zero temperature distribution, these properties will 
be independent of $q$! Of course, the SOS configurations 
allowed for $q=2$ are very different from the froth that 
appears for $q>2$, and thus in principle, we expect two 
different universality classes. Strictly speaking, since 
four colors are needed to cover an arbitrary ``map" in 
two dimensions, there are restrictions on configurations 
allowed for $q=3$. If relevant, this leads to a third 
potential random interface behavior in $d=2$. Indeed the 
recursion relations of the DDHL at $T=0$ discriminate 
between $q=2$, $q=3$, and $q\geq4$.
 
We focused our studies of critical behavior at $T=0$ on 
the mixture of bonds of strengths $+J$ and $-J$. The 
critical concentration in the presence of bubbles was 
identified by trial and error as $p_c=0.883\pm0.001$. 
At the critical point, the mean and variance of energy 
should scale as $\overline{E(L)}=L^{\theta_c}$ 
and ${\rm var}[E(L)]=L^{2\omega_c}$, with 
$\theta_c=\omega_c$ if there is only one energy scale. 
The fits at $p_c$ are consistent with this expectation, 
giving $\theta_c=0.34\pm 0.05$ and $\omega_c=0.31\pm0.08$.
Rather surprisingly, $\omega_c$ is very close to the value 
of $\omega$ for directed paths with no loops. Fig.(2) 
shows that the data for the energy on approaching $p_c$ 
from above can be collapsed by a finite--size scaling form,
\begin{equation}\label{CollapseE}
E(L,p\geq p_c)=A|p_c-p|^\mu L+BL^{\omega_c},
\end{equation}
with $\mu=0.91\pm0.04$. In the absence of loops for $q=2$, 
$\mu=1$ exactly, since the mean value of the final energy 
is simply linear in the mean value for individual bonds. 
There is also a subleading correction to the energy from 
fluctuations that scales as $L^{\omega_c}$\cite{Orland}.

To better understand the closeness of exponents in the 
presence and absence of loops, we looked directly at the 
fractal structure of the critical interface. This is 
achieved by examining the probability $P_n$ that a loop 
forms at the $n^{\rm th}$ iteration. As indicated in 
Fig.(3), $P_n$ decays exponentially to zero for $p>p_c$,
while the decay slows down on approaching $p_c$. The data 
at $p_c$ can be fitted to a decay as $c/n$, with $c=0.060\pm 0.003$. 
This fit suggests the {\it phenomenological} differential 
recursion relations
\begin{equation}\label{DRR}
{dP \over dn}=-f_sP-P^2/c,\qquad
{df_s \over dn}=f_s+f_sP .
\end{equation}
The first equation reproduces $P(n)=c/n$ at criticality; 
the second is just the behavior of the mean (free) energy 
(or mass of the cluster). The fixed point at $f_s=P=0$
describes a critical cluster that is asymptotically linear, 
with loops appearing 
predominantly at short length scales. Linearizing the 
second equation gives $y=1$. However, the marginality 
of $P$ at criticality leads to logarithmic corrections 
to various scaling quantities. For example, the mass of 
the critical cluster grows as $L(\ln L)^c$. Similarly,
the interfacial free energy vanishes as
$|p-p_c||\ln(|p-p_c|)|^c$. Fig.(2) shows that a data
collapse is also possible using such logarithmic 
corrections. The best fit is achieved for $c=0.5\pm 0.6$, 
compared to $c=0.060\pm 0.003$ obtained directly from Fig.(3).

The asymptotic linearity of the critical cluster extends 
beyond the simple example of the DHL, and was also checked for 
the DDHL. In the latter, an interface configuration that covers 
both branches, as well as the central diagonal, is allowed
only for $q>3$. Thus this lattice supports three types of 
recursion relations at $T=0$, corresponding to $q=2$, $3$, 
and $\geq4$. In spite of these differences, within our 
numerical accuracy, we could not detect any significant changes 
in the exponents from the simpler DHL. As indicated in table I, 
even in the absence of loops ($q=2$), the critical cluster in 
the pure system is a fractal. This is because it takes advantage 
of the diagonal bond and is no longer simply directed. The 
critical cluster of the random system does not take advantage 
of the diagonal bond, or the possibility of loops, staying 
asymptotically linear. This suggests that the marginal irrelevance 
of operators that may complexify the structure (such as $P$ 
in eq.(\ref{DRR})) is in fact quite generic. 
%Such marginal irrelevance has also been observed for potentials 
%that pin the interface\cite{Balents}.

Although we have presented the results in the context of Potts 
models, they are probably more generally applicable to systems 
of discrete symmetry. For example, it has been suggested that 
regions of the Baxter model with diverging heat capacity also 
exhibit Ising like behavior in the presence of random bonds
\cite{Matthews}\cite{Dotsenko}. It is also tempting to generalize the 
conclusions to higher dimensions: Strong enough disorder may 
result in a continuous transition\cite{Berker} with the 
interfacial criticality governed by a zero temperature
fixed point. If the apparent irrelevance of bubbles can be 
generalized from  hierarchical lattices, the exponent $\mu$ 
will be the same as in the Ising model. 
How can we reconcile this apparent super--universality
of random bond criticality, with earlier results\cite{RBRG}\cite{Ludwig}
which do indicate exponents that depend on $q$? The latter
calculations were performed for weak disorder, and lead to
a finite temperature fixed distribution. Is it possible that
stronger disorder leads to different behavior, dominated by
zero temperature fixed points? This scenario is precisely
what is observed in a recent position space RG of a three
state random bond model\cite{Falicov}. Of course, another
possibility is that our approximate interface model does
not fully capture the physics of the system. Clearly, 
further investigations are desirable.

In summary, we have introduced a simple position space 
approximation for studying the interfacial properties of Potts 
models in $d=2$. In the pure case, the critical interface is 
a fractal froth, and interfacial tension vanishes with an 
exponent $\mu(q)$. In a random medium,  within our numerical 
accuracy, we find that the critical interface is asymptotically
linear, and the interfacial free energy vanishes linearly with 
logarithmic corrections, i.e. $\mu=1$ independent of $q$. We 
argue that, {\it if governed by a zero  temperature fixed 
point}, $\mu(q)$ should be independent of $q$ for all 
sufficiently  large $q$. The above model also provides the 
simplest generalization of directed  paths in random 
media\cite{DP} to ramified objects, with potential applications to fracture cracks, lightning patterns, etc. 
%Various extensions and improvements of the position space RG 
%are possible, and currently under study. 
%Details will be given elsewhere.

We acknowledge discussions with N.C. Bartelt, A.N. Berker, 
and T.L. Einstein. This research was initiated
during visits to Oxford University (MK and AS, who also 
acknowledges an Accademia dei Lincei- Royal Society grant 
and INFM support), and the Isaac Newton Institute (BD and MK). 
The work at MIT was supported by the NSF through grants 
DMR-93-03667 and PYI/DMR-89-58061.

\begin{table}
\label{Exponents}
\caption{The exponent $\mu(q)$ for the vanishing of the interfacial free energy,
obtained from the DHL and  the DDHL, compared to the exact  values in $d=2$.}
\begin{tabular}{cccc}
$\quad q\quad$&DHL&DDHL&\, Exact \cr
\hline
1&1.634&1.429&4/3\cr
2&1&0.847&1\cr
3&0.886&0.749&5/6\cr
4&0.830&0.700&2/3\cr
\end{tabular}
\end{table}

\begin{figure}
\vskip 5cm
\caption{Phase boundaries for $q=3$ and 4, with mixtures of 
positive and negative bonds (solid, dashed), and positive bonds (dotted, dashed dotted).
Arrows indicate RG flow towards the $T=0$ fixed point.
The solid lines in the inset figure indicate the cell replacing
each bond at every stage of the DHL construction.  
The central (dashed) diagonal is also
present in the DDHL.}
\end{figure}

\begin{figure}
\vskip 5cm
\caption{Data collapse of $E(L,p\geq p_c)/L^{\omega_c}$,
against $(p-p_c)^\mu L^{(1-\omega_c)}$ (left group), 
or $|p-p_c||\ln(|p-p_c|)|^c L^{(1-\omega_c)}$ (right group) 
in the abscissa. The scales are logarithmic.}
\end{figure}

\begin{figure}
\vskip 5cm
\caption{The iterated probability for loop formation $P_n$.
The decay is fitted to $c/n$ (asterisks and dashed line) 
at $p_c$. Data for $p>p_c$ (squares) can be fitted 
to an exponential.}
\end{figure}

%\end{multicols}

\begin{references}
\bibitem{Dist} The initial ensemble is consistent with 
${\cal P}_1$. At each iteration, groups of four bonds are 
sampled and from each group a new weight is obtained through 
eq.(\ref{ZRR}).
\end{references}
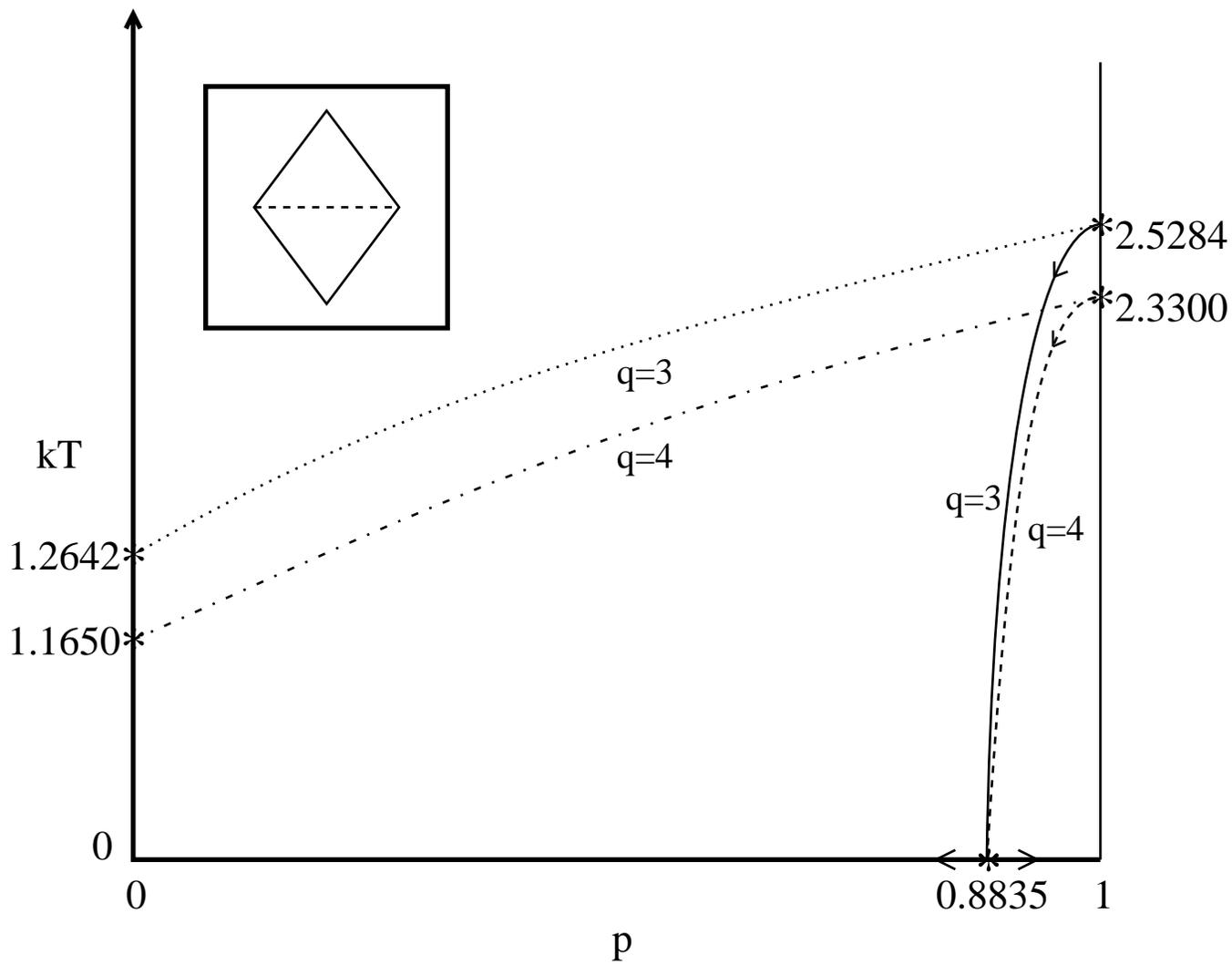
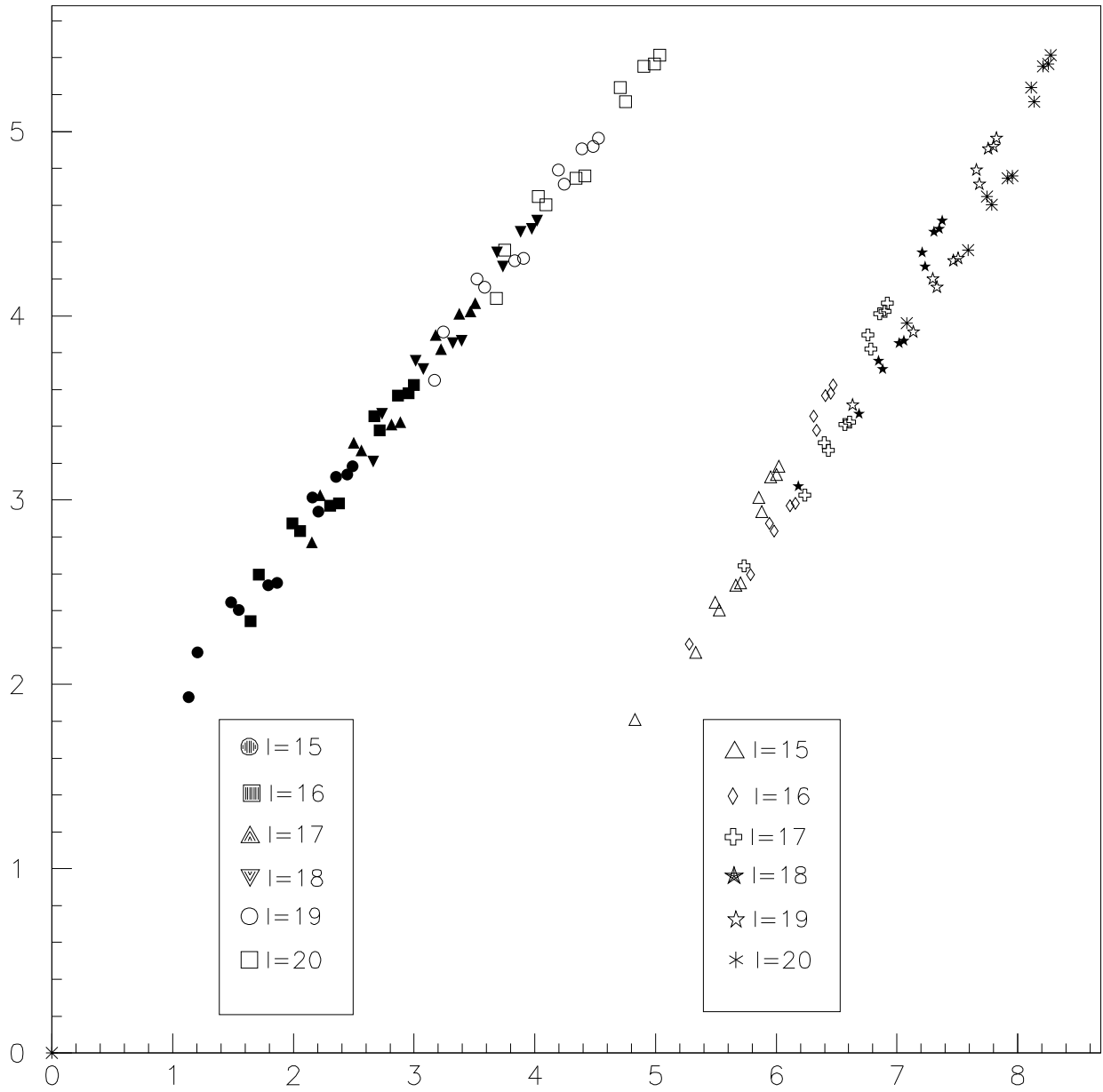
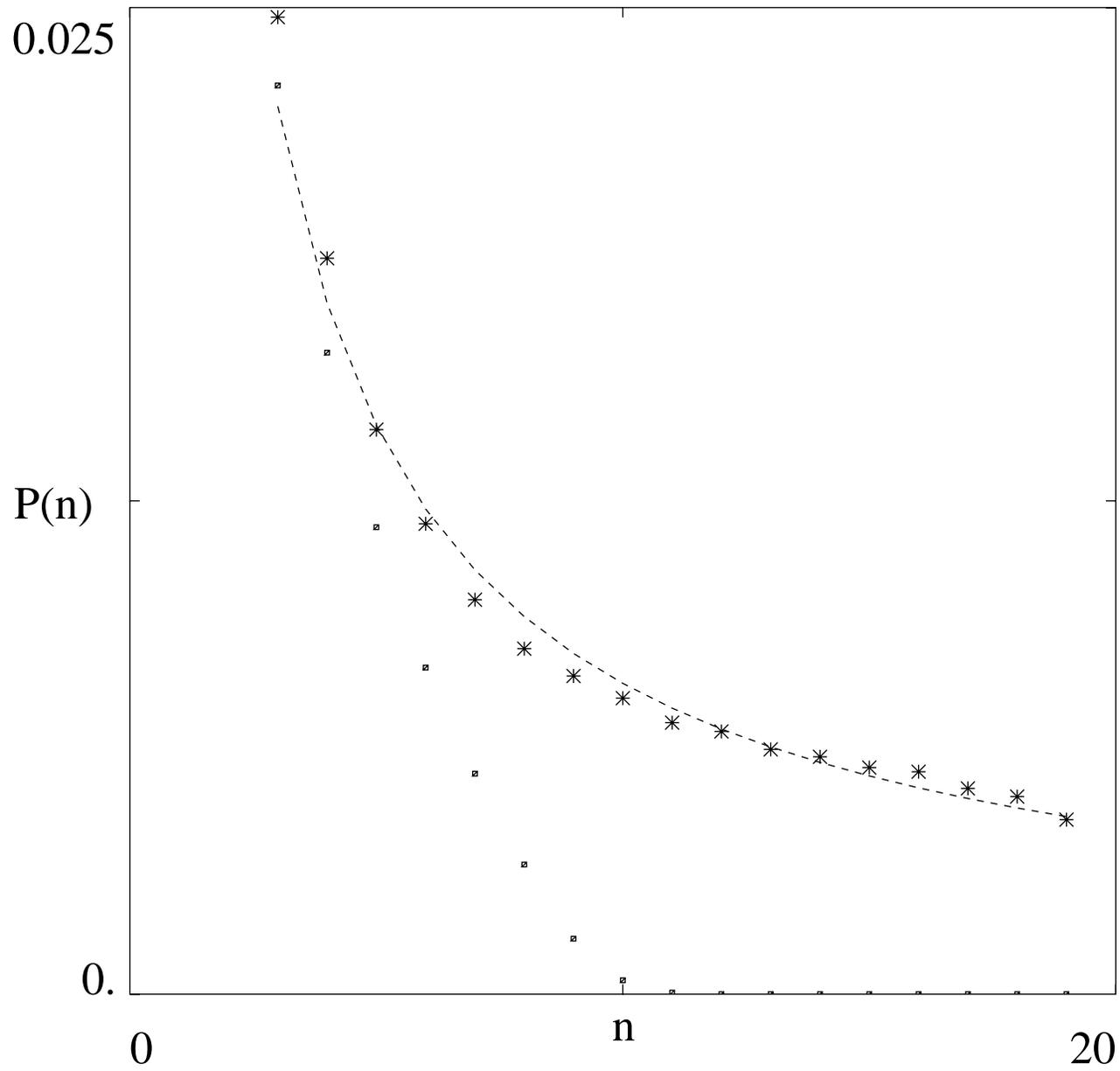
\end{document}